\documentclass[sigconf,natbib=true,anonymous=false,review=false]{acmart}

\AtBeginDocument{%
  \providecommand\BibTeX{{%
    \normalfont B\kern-0.5{\scshape i\kern-0.25 b}\kern-0.8\TeX}}}

\usepackage{natbib}

\usepackage{booktabs} 
\usepackage{tabularx} 
\usepackage[english]{babel}

\usepackage{siunitx}  

\usepackage{multirow}
\usepackage{multicol}
\usepackage{array}
\usepackage{arydshln}
\usepackage[flushleft]{threeparttable}

\usepackage{enumitem}

\usepackage{caption}
\usepackage{subfigure}

\usepackage{footnote}
\makesavenoteenv{tabular}
\makesavenoteenv{table}
\usepackage{bbm}
\usepackage{algorithm}
\usepackage{algpseudocode}

\usepackage{xspace} 

\usepackage{hyphenat}

\usepackage{pifont}
\newcommand{\xmark}{\text{\ding{55}}}
\newcommand{\ymark}{\text{\ding{52}}}

\copyrightyear{2023}
\acmYear{2023}
\setcopyright{acmlicensed}\acmConference[ICTIR '23]{Proceedings of the 2023
ACM SIGIR International Conference on the Theory of Information Retrieval}{July
23, 2023}{Taipei, Taiwan}
\acmBooktitle{Proceedings of the 2023 ACM SIGIR International Conference on
the Theory of Information Retrieval (ICTIR '23), July 23, 2023, Taipei, Taiwan}
\acmPrice{15.00}
\acmDOI{10.1145/3578337.3605127}
\acmISBN{979-8-4007-0073-6/23/07}

\begin{document}

\title{Dense Retrieval Adaptation using Target Domain Description}

\author{Helia Hashemi}
\orcid{0000-0001-7258-7849}
\authornote{Part of this work was done during a research internship with Bloomberg.}
\affiliation{
  \institution{University of Massachusetts Amherst}
  \country{United States}
}
\email{hhashemi@cs.umass.edu}

\author{Yong Zhuang}
\orcid{0000-0002-7858-5569}
\affiliation{
  \institution{Bloomberg}
  \country{United States}
}
\email{yzhuang52@bloomberg.net}

\author{Sachith Sri Ram	Kothur}
\orcid{0009-0006-4858-3737}
\affiliation{
  \institution{Bloomberg}
  \country{United States}
}
\email{skothur@bloomberg.net}

\author{Srivas Prasad}
\orcid{0009-0006-0379-7416}
\affiliation{
  \institution{Bloomberg}
  \country{Canada}
}
\email{sprasad60@bloomberg.net}

\author{Edgar Meij}
\orcid{0000-0003-0516-3688}
\affiliation{
  \institution{Bloomberg}
  \country{United Kindgom}
}
\email{emeij@bloomberg.net}

\author{W. Bruce Croft}
\orcid{0000-0003-2391-9629}
\affiliation{
  \institution{University of Massachusetts Amherst}
  \country{United States}
}
\email{croft@cs.umass.edu}

\renewcommand{\shortauthors}{Helia Hashemi et al.}


\begin{abstract}

In information retrieval (IR), domain adaptation is the process of adapting a retrieval model to a new domain whose data distribution is different from the source domain. Existing methods in this area focus on unsupervised domain adaptation where they have access to the target document collection or supervised (often few-shot) domain adaptation where they additionally have access to (limited) labeled data in the target domain. There also exists research on improving zero-shot performance of retrieval models with no adaptation. This paper introduces a new category of domain adaptation in IR that is as-yet unexplored. Here, similar to the zero-shot setting, we assume the retrieval model does not have access to the target document collection. In contrast, it does have access to a brief textual description that explains the target domain. We define a taxonomy of domain attributes in retrieval tasks to understand different properties of a source domain that can be adapted to a target domain. We introduce a novel automatic data construction pipeline that produces a synthetic document collection, query set, and pseudo relevance labels, given a textual domain description. Extensive experiments on five diverse target domains show that adapting dense retrieval models using the constructed synthetic data leads to effective retrieval performance on the target domain. 

\end{abstract}



\maketitle

\section{Introduction}
\label{sec:intro}

The effectiveness of neural information retrieval (IR) models has been well-established in recent years \cite{trec2020overview,Guo:2020,Mitra:2018}. However, these models have primarily demonstrated strong performance in settings where the training and test data follow a similar data distribution \cite{thakur:2021}. When well-performing neural models developed for one test collection, e.g., MS MARCO \cite{MSMARCO}, are applied to a substantially different one, the results are often worse than those produced by much simpler bag-of-words models such as BM25 \cite{Robertson:1995}. This poses a problem in real-world applications, where access to large, domain-specific training data is limited. To address this issue, a group of methods known as ``domain adaptation'' have been developed.

There are various approaches to domain adaptation in information retrieval, as summarized in Table \ref{tab:domain-setting}. In the zero-shot setting, the assumption is that the model has been trained on a large-scale test collection in a source domain, but no data from the target domain is available during training. It is worth noting that in the zero-shot setting, there is essentially no adaptation taking place, as the model is simply being tested on the target domain. In contrast, unsupervised domain adaptation models assume that the target document collection is available for adaptation. The few-shot setting takes this further and assumes that a small set of query-document pairs with relevance labels on the target domain is available, allowing the retrieval model to be adapted to the target.

In this work we introduce a new category of domain adaptation methods for neural information retrieval, which we refer to as ``domain adaptation with description.'' Studying this problem is not only interesting from an academic perspective, but also has potential applications in several real-world scenarios, where the target collection and its relevance labels are not available at training time. For example, these may not be available yet or at all or, even if they were, target domain owners may be hesitant to provide them for various reasons such as legal restrictions. There are also applications with privacy concerns, for instance in the case of medical records or where the data contains personally identifiable information. Another example can be found when a competitive advantage is involved, as potential use of the data may benefit competitors. Therefore, if an organization lacks the resources for training neural IR models in-house and desires to outsource the process, they should be able to provide a high-level textual description that outlines the task and characteristics of the data in a general manner. Our approach then allows the organization to convey the necessary information to a third party without compromising sensitive information or violating legal restrictions.

\begin{table*}[t]
    \centering
    \vspace{-0.3cm}
    \caption{Different categories of domain adaptation in information retrieval.}
    \vspace{-0.3cm}
    \resizebox{\textwidth}{!}{
    \begin{threeparttable}
        \begin{tabular}{lcccl} 
        \toprule
        \textbf{Adaptive Retrieval Setting} & \textbf{$q$-$d$-$r$ triplets$^\dagger$ in $D_1$} & \textbf{$q$-$d$-$r$ triplets in $D_2$} & \textbf{Target Collection} & \textbf{Extra Information} \\\midrule
        Zero-shot retrieval & \textcolor{green}{\ymark} & \textcolor{red}{\xmark} & \textcolor{red}{\xmark} & None \\
        Unsupervised domain adaptation & \textcolor{green}{\ymark}  & \textcolor{red}{\xmark} & \textcolor{green}{\ymark} & None \\
        Supervised (few-shot) domain adaptation & \textcolor{green}{\ymark}  & \textcolor{green}{\ymark}$^\ddagger$ & \textcolor{green}{\ymark} & None \\\hdashline
        \textbf{Domain adaptation with description} & \textcolor{green}{\ymark}  & \textcolor{red}{\xmark} & \textcolor{red}{\xmark} & a textual description of the target domain$^*$ \\
        \bottomrule
        \end{tabular}
        \begin{tablenotes}
          \scriptsize
          \item $\dagger$ $q-d-r$ refers to training data triplets of query, document, and relevance labels. \qquad \qquad  $\ddagger$ often only a small amount of training data is available. \qquad \qquad  $*$ domain description can be a single sentence describing the target domain.
      \end{tablenotes}
    \end{threeparttable}
    }
    \vspace{-0.3cm}
    \label{tab:domain-setting}
\end{table*}

In this paper, we investigate the task of domain adaptation for information retrieval (IR) tasks by utilizing target domain descriptions. We propose a taxonomy for the task and analyze the various ways and attributes by which a domain can be adapted. We differentiate our task from similar studies that have been conducted in recent years and explain the limitations of existing technologies. To address these limitations, we propose a novel pipeline that utilizes the domain descriptions to construct a synthetic target collection and generate queries and pseudo relevance labels to adapt the initial ranking model trained on a source domain. Our approach takes advantage of state-of-the-art instruction-based language models to extract the properties of the target domain based on its given textual description. We show that a retrieval-augmented approach for domain description understanding can effectively identify various properties of each target domain, including the topic of documents, their linguistic attributes, their source, etc. The extracted properties are used to generate a seed document using generative language models and then an iterative retrieval process is employed to construct a synthetic target collection, automatically. 

Following prior work on unsupervised domain adaptation \cite{wang:2021}, we automatically generate queries from our synthetic collection based on the query properties extracted from the target domain description. We then generate pseudo relevance labels for each query given an existing cross-encoder reranking model and use the created data for adapting \textit{dense retrieval} models to the target domain. 
Extensive experiments on five diverse target collections, ranging from financial question answering to argument retrieval for online debate forums, demonstrate the effectiveness of the proposed approach for the task of domain adaptation with description. In summary, the main contributions of this work include the following.
\begin{itemize}[leftmargin=*]
    \item Introducing the novel task of domain adaptation with description for information retrieval.
    \item Proposing an automatic data construction pipeline from each target domain description.
    \item Proposing a taxonomy of domain attributes in information retrieval that should be identified for effective adaptation.
    \item Introducing an effective implementation of the proposed pipeline.
\end{itemize}

\vspace{-0.3cm}
\section{Related Work}
This work is related to the domain adaptation as well as prompt-based language model literature. 

\subsection{Domain Adaptation in Neural IR}
Research in this area can be categorized into two main groups: supervised and unsupervised.  In supervised (often few-shot) domain adaptation, the assumption is that labeled data is available in the source domain and a limited amount of labeled data is available in the target domain. This problem can be formulated as a few-shot learning scenario, as demonstrated by \citet{sun:2020}. A common approach within this category is transfer learning, which utilizes a pre-trained model from the source domain and fine-tunes it on the target domain using a small set of labeled data. This approach has been shown to improve model performance by allowing the model to learn the specific characteristics of the target domain \cite{devlin:2018}.

The unsupervised setting assumes that access to target documents is available, but queries and relevance labels are not. \citet{wang:2021} proposed a generative pseudo-labeling approach for this scenario. They generated synthetic queries, from documents and applied a re-ranking based pseudo labeling approach for each query and document pair. Then, the model was fine-tuned using the generated query-document pairs. 
\citet{zhu:2022} proposed an answer-aware strategy for domain data selection, which selects data with the highest similarity to the new domain. The source data examples were sorted based on their distance to the target domain center, and the most similar examples were chosen as pseudo in-domain data to re-train the question generation model. Additionally, they presented two confidence modeling methods, namely, generated question perplexity and BERT fluency score, which emphasized labels that the question generation model was more confident about. Recently, \citet{gao:2022} introduced a zero-shot dense retrieval model for adaptations by using a generative model to generate hypothetical documents relevant to the query. These documents were used as queries and, with the use of pre-trained Contriever \cite{izacard:2021}, documents from the target domain were retrieved.

\vspace{-0.3cm}

\subsection{Prompt-based Language Models}
\label{sec:rel-lm}

Language models have been widely used in information retrieval (IR) and natural language processing (NLP) applications due to their ability to accurately represent text. They are machine learning models that are trained to predict the likelihood of a sequence of words. 
Currently, the state-of-the-art approach is to use large transformer-based language models, such as BERT \cite{devlin:2018}, GPT \cite{radford:2018}, and T5  \cite{Raffel:2019}. An evolving technique for training these models is called ``prompting.'' GPT-3 \cite{brown:2020} is an example of a successful language model that was trained using this technique. Prompting refers to using language models to generate text by providing the model with a ``prompt,'' which is a short text that serves as a starting point for the model's generation. The idea behind prompting is to provide the model with a specific context or task, so that it can generate text that is more focused and coherent.


Prompts can be used for few-shot learning. To be more specific, language models can be fine-tuned for specific tasks using a small amount of task-specific data, such as a few examples or instructions. These type of models are called instruction-tuned language models. They include T0 \cite{sanh:2021}, InstructGPT \cite{ouyang:2022}, and \textsc{T$k$-Instruct} \cite{wang:2022}. Instruction-tuned models are promising in that they make it possible to fine-tune language models on new tasks with minimal data. InstructGPT \cite{ouyang:2022} argues that it is more effective and truthful than GPT-3 at following user intention. 

In this context, the term ``instruction'' is distinct from ``description'' as used in this paper. In previous research, the term ``instruction'' has been used interchangeably with ``intention'' and is closely related to the concept of user intent in the field of IR. For example, it was found that if GPT-3 prompted to explain the moon landing to a 6-year-old, it outputs the completion of the prompt text, while InstructGPT generates a more accurate and appropriate response that actually explains moon landing with simple wording \cite{openai-moon}. This is attributed to their training -- GPT-3 predicts the next word, while InstructGPT  employs techniques such as reinforcement learning from human feedback for fine-tuning the model to better align with user instructions. Other recent research has focused on fine-tuning language models to follow instructions using academic NLP datasets such as FLAN \cite{wei:2021} and T0 \cite{sanh:2021}. However, all these instruction-based language models are currently limited in their ability to perform complex, multi-step tasks, as opposed to the high-level task-oriented approach used in this study.

Instruction-tuned language models have been effectively applied to various NLP tasks, but have received less attention in the field of IR. This is due to the challenge of casting a retrieval task into the sequence-to-sequence format typically used by these models, as it requires encoding a large corpus of documents.
Concurrent to our work \citet{asai:2022} proposed a retrieval method that explicitly models a user's search intent by providing natural language instruction. They concatenated the query with the instruction, encoding it as the query embedding, and then computed the cosine similarity between query and document pairs. \citet{gao:2022} used InstructGPT to encode a query with its instruction and generated a hypothetical document, which they later used as the query to improve dense retrieval. While we use both of these ideas in our baselines, our approach in defining the task differs, significantly. In both of the aforementioned papers, the authors simply concatenated the instruction to the query. However, this approach is limited to handling atomic commands that improve alignment with human intentions, such as ``write an answer to this question.'' These types of instructions are distinct from high-level overviews of complex tasks that require multiple steps to complete, such as our task.

\begin{figure*}
    \centering
    \vspace{-0.3cm}
    \includegraphics[trim="2cm 4.5cm 3.5cm 4cm",clip,width=.68\textwidth]{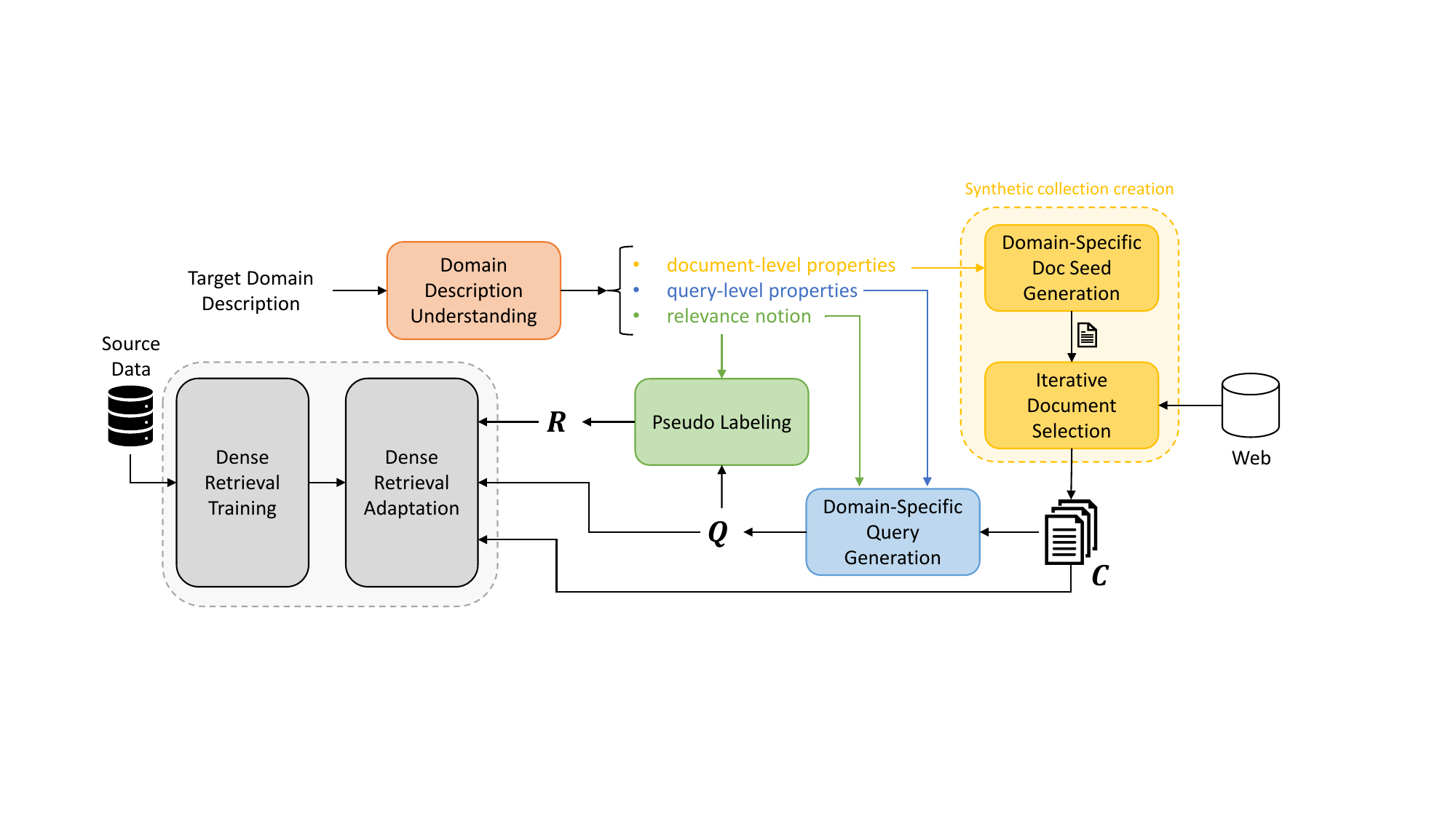}
    \vspace{-0.3cm}
    \caption{The proposed pipeline for training dense retrieval models for a given domain description.}
    \label{fig:pipeline}
    \vspace{-0.3cm}
\end{figure*}

\section{Methodology}

 In this section, we explain the problem formulation and a taxonomy of domain attributes that can be used to understand domain descriptions. Such domain understanding component can produce attribute values for a synthetic corpus construction model that uses a large language model to generate one seed document with these attributes and then performs an iterative retrieval process from a heterogeneous collection such as the Web for collection creation. The constructed collection will be then used to generate queries and pseudo relevance labels that are aligned with the properties of the target domain, as extracted by our domain understanding component. This pipeline leads to a synthetic training set that can be used to adapt a dense retrieval model to the target domain.

\subsection{Problem Formalization}
\label{sec:formal}
Let $M$ be a retrieval model that is trained on the source domain $D_1$, and $T$ be the textual description of the target retrieval domain $D_2$, where $D_2 \neq D_1$. The goal is to adapt the retrieval model $M$ to the target domain $D_2$ and obtain the retrieval model $M'$ that performs effectively on $D_2$. Assume that $W$ is a large-scale  heterogeneous collection, such as a Web collection, that can be used as an external resource as required. This large-scale collection can be used for synthetic collection construction for any target domain description. 

\begin{table*}[t]
    \centering
    \small
    \caption{A taxonomy of attributes that define an information retrieval task.}
    \vspace{-0.3cm}
    \resizebox{\textwidth}{!}{
    \begin{threeparttable}
    \begin{tabular}{clp{7cm}p{7cm}}\toprule
        & \textbf{Retrieval Attribute} & \textbf{Attribute Definition} & \textbf{Example Attribute Values} \\ \midrule
        \multirow{7}{*}{\rotatebox[origin=c]{90}{Query Attributes}} & Query topics\textsuperscript{*} & the subject matters or themes of the users' search requests &  medical, financial, climate, etc. \\
        & Query linguistic features  & syntactic characteristics of the query &  formal, informal, technical, etc.\\
        & Query language &  a language used by the user to make requests for information& English, Spanish, etc.   \\
        & Query structure & the structure of the query used by the user & structured, semi-structured, unstructured, SQL, etc.  \\
        & Query modality & the query modality  & text, text and image, uni-modal, multi-modal, etc. \\
        & Query format & type of the query submitted by the user & keyword queries, tail queries, tip-of-tongue queries, etc. \\
        & Query context & any metadata that exists around the query  & conversational search, session search, from adult users vs. kids  \\\hdashline
        \multirow{7}{*}{\rotatebox[origin=c]{90}{Doc Attributes}} & Document topics &  the main subjects that the document collection cover & medical, financial, etc.  \\
        & Document linguistic features &  syntactic characteristics of the documents & formal, informal, technical, etc. \\
        & Document language & the language used to express the content of the documents & English, Spanish, etc. \\
        & Document structure  & the structure of the documents in the collection & structured, semi-structured, unstructured, knowledge base, etc. \\
        & Document modality & the document modality &  text, text and image, uni-modal, multi-modal, etc. \\
        & Document format &  the format of the document (especially from IR perspective) & passages, long documents, questions, etc.   \\
        & Document source &  the specific source that the documents come from & Wikipedia, Twitter, Quora, etc.   \\ \hdashline
        & Relevance notion  &  the criteria that make the documents relevant to the query & topical relevance, containing the correct answer, paraphrasing, containing the counterargument, etc. \\
         \bottomrule
    \end{tabular}
          \begin{tablenotes}
          \scriptsize
        \item[*] This is often referred to as the ``domain'', but we use the term ``topic'' to avoid confusion.
      \end{tablenotes}
    \end{threeparttable}
    }
    \label{tab:taxonomy}
    \vspace{-0.3cm}
\end{table*}

\subsection{A Taxonomy of Domain Attributes in IR}

 The term ``domain'' is used quite loosely in NLP and IR and defined in myriad ways
 \cite{plank:2016}. It is commonly used to describe a type of corpus that is ``coherent'', such as a specific topic or linguistic register \cite{plank:2011}. However, the concept of domain has evolved in recent years, leading to ongoing research in this area. For example, there is a distinction between ``canonical'' data (e.g., edited news articles) and ``non-canonical'' data (e.g., social media), and models trained on one type may not perform well on the other. There is an ongoing debate over what constitutes a ``domain'' in the field of information retrieval (IR), and whether subdomains exist within a larger domain. This uncertainty makes it difficult to tackle the domain adaptation problem and develop a universal algorithm, as domain shifts are specific to each case and models may not perform robustly when transferred from one case to another. 
 
 In order to clarify the different stances on the definition of a ``domain'' we have developed a taxonomy for domains and their attributes in the context of IR. Therefore, we define a domain based on the set of attributes defined in our taxonomy. This taxonomy can be used to develop general-purpose domain adaptation solutions as it enumerates the possible ways in which two domains can be different. We argue that every retrieval task is composed of three variables: query, documents, and relevance notion. We propose that attributes related to these three categories together define a retrieval domain. In other words, for any domain $D$, we define a set of attributes $\{a_1, a_2, \cdots, a_n\}$, where each attribute $a_i$  is either related to the properties of query, document, or relevance. Through careful exploration of many different retrieval tasks, including the ones in the BEIR benchmark \cite{thakur:2021} and the ones organized by TREC and CLEF evaluation campaigns over the last few decades, we compile a taxonomy that includes seven query-level attributes, seven document-level attributes, and one attribute denoting the relevance notion. The attributes, their definition, and examples are presented in Table \ref{tab:taxonomy}. In the interest of space, do not list them here again. We argue that if the value of at least one attribute belonging to any of the three categories changes, a domain shift has occurred. We highlight the asymmetric nature of query and document attributes that presents unique challenges for domain adaptation in IR compared to NLP tasks. Finally, we note this taxonomy can be used to see what attributes differ between domains and that we can leverage those for effective  adaptation.

\subsection{Domain Description Understanding}
\label{sec:desc-und}

 As discussed in Section \ref{sec:intro}, clients may be reluctant to provide actual target domain data. However, providing a high-level description of the data is usually feasible. At the time of this research, no dataset that includes descriptions of retrieval tasks were known to us. Concurrently, \cite{asai:2022, gao:2022} provided instructions for some IR test collections. However, we started this research prior to their work being submitted to arXiv (Dec 2022) and as noted in Section \ref{sec:rel-lm}, the instructions they use provide more fine-grained information on human intentions, in line with what was referred to as ``narratives'' in the TREC 2004 Robust Track \cite{voorhees:2004}. That being said, in our problem, we need a description of the \textit{retrieval task} that includes information on the appearance of the corpus and queries, in addition to user intentions, and how relevance is defined for that task. To obtain these descriptions, we gave 15 diverse IR collections from the BEIR benchmark \cite{thakur:2021} to three IR experts (not the authors of this paper) and asked them to explain the retrieval task for each. We asked them to revise the differences of opinion during a brainstorming session; they shared their explanations and worked together to reach a single description for each collection, which we refer to as $T$ in our formalization. After the descriptions are finalized, we provide the same people with the taxonomy we have defined in Table \ref{tab:taxonomy}, and ask them to annotate the descriptions based on the taxonomy attribute. This annotation results in the gold labels of attribute values based on our taxonomy for each dataset. We provide one dataset description and its annotation in Table \ref{tab:annotation-example} for the reference.

We argue that a proper understanding of the description has a significant impact on adaptation. If the model understands the value of each attribute in the taxonomy, it knows when a domain shift has occurred and what attributes need to be adapted for the entire model to be adapted. 
Therefore, our domain description understanding component focuses on predicting the values of attributes defined in our taxonomy. Since the value of the attributes can be open-ended text rather than defined options, the best architectural choice is a text generation model that takes the domain description as input and generates the value of the attributes as output. Therefore, we adopt a state-of-the-art prompt-based text generation model $F$ to perform the task, i.e., ChatGPT. We instruct the model to get the description of the domain and extracts the value of attributes introduced in the taxonomy.\footnote{After some rounds of trial and error, we landed on the following instruction, $I$ as the best performing one for our task:
``For each defined retrieval task in the Passage, find the values related to the relevance notion (e.g., topically relevant, contains the answer, references of a paper, paraphrase, evidence for the claim, etc.) as well as the following query and document attributes: query topic (e.g., medical, scientific, financial, mathematical, adult, etc.); query linguistic features (e.g., formal, informal, etc.); query language (e.g., english, french, etc.); query structure (e.g., unstructured, semi-structured, structured, etc.); query modality (e.g., text, image, video, etc.); query format (e.g., keyword query, tail query, question, claim, argument, passage, etc.); document topic (e.g., medical, scientific, financial, mathematical, adult, etc.); document linguistic features (e.g., formal, informal, etc.); document language (e.g., english, french, etc.); document structure (e.g., unstructured, semi-structured, structured, etc.); document modality (e.g., text, image, video, etc.); document format (e.g., passage, long document, question, etc.); document source (e.g., StackExchange, wikipedia, reddit, youtube, twitter, facebook, quora, etc.).
If the value of each attribute cannot be inferred, return NA''}

\begin{table*}[t]
    \centering
    \caption{An example of a  retrieval task description and its annotated attributes from our taxonomy.}
    \vspace{-0.3cm}
    \resizebox{\textwidth}{!}{
        \begin{tabular}{p{3cm}p{15cm}} 
        \toprule
        \textbf{Target Collection} & Arguana \\\midrule
        \textbf{Description of the retrieval task} & Given an argument passage as a query, the task is to retrieve passages from online debate portals that contain its counterarguments  \\\midrule
        \textbf{Description annotation} & relevance notion: counterargument $\blacksquare$ query topic: NA $\blacksquare$ query linguistic features: NA $\blacksquare$ query language: NA $\blacksquare$ query structure: unstructured $\blacksquare$ query modality: unimodal $\blacksquare$ query format: argument passage $\blacksquare$ document topic: NA $\blacksquare$ document linguistic features: NA $\blacksquare$ document language: NA $\blacksquare$ document structure: unstructured $\blacksquare$ document modality: unimodal $\blacksquare$ document format: argument passage $\blacksquare$ document source: online debate portals \\\bottomrule
        \end{tabular}
    }
    \vspace{-0.3cm}
    \label{tab:annotation-example}
\end{table*}

In addition to the instruction, we include up to three examples from the most similar collections to the target domain by retrieval augmentation. Let $R(T,C')$ denote a retrieval model (SBERT in our case) that takes the target domain description and a collection of textual descriptions of different domains ($C'$). 
The description understanding function $F$ takes the instruction $I$ , the retrieved examples, and domain description $T$, and outputs the values of attributes introduced in taxonomy. Formally: $F(I, T, R(T, C')) = \{a'_1, a'_2, \cdots, a'_n\}$ where $n=15$.

\noindent \paragraph{\textbf{Discussion}} One may argue that the taxonomy is easy to understand and interpret, therefore, users can directly identify these properties for the target domain and this bypasses the need to a Domain Description Understanding component. This argument is valid. In other words, the taxonomy we define in Table~\ref{tab:taxonomy} enables users of the system to directly identify the values of each attributes for the target domain. That being said, the Domain Description Understanding component enables the users to just describe their target domain in natural language. Similar to any semantic parsing task, such as text-to-SQL, this component creates a natural language interface for this task. Thus, studying it can shed light into how feasible it is to extract domain attributes from natural language.




\subsection{Synthetic Target Data Construction}
\label{sec:method}
As depicted in Figure~\ref{fig:pipeline}, once we identify the domain attributes of our taxonomy  for the target domain (i.e., domain description understanding), we propose to build a synthetic training set based on the generated attribute values. This consists of three steps: synthetic document collection construction, synthetic query generation, and pseudo-labeling. In the following we describe each of these steps. Our data construction approach is presented in Algorithm \ref{alg:greedy-cross}.


\subsubsection{\textbf{Synthetic Document Collection Construction}}
One naive approach to synthesizing the collection is to generate documents one by one using sequence\hyp{}to\hyp{}sequence models. In preliminary experiments, we observed that many state-of-the-art and free\hyp{}to\hyp{}use sequence\hyp{}to\hyp{}sequence models such as the latest version of \textsc{T$k$\hyp{}Instruct} \cite{wang:2022}, are not sufficient to generate meaningful documents given our target domain descriptions. Instead, they generate passages containing words from our instructions, rather than generating a document with the provided attributes.

It can be argued that with the rise of black-box generative language models like ChatGPT, this issue will be reduced. However, it is important to note that these models are not free to use. At the time of submitting this paper, ChatGPT was not yet available through an API, so we used the next best available large language model, \texttt{text-davinci-003}, the latest version of GPT-3 from OpenAI. OpenAI charges customers based on the cumulative number of tokens in the input and output, at a rate of \$0.02 per 1K tokens. If we consider an average passage to be 300 tokens, the minimum cost to generate a corpus like MS MARCO (consisting of 8M passages) would be \$12,000. This assumes the model only takes the domain description with no example as input and generates one passage in line with the target domain description.
 
It is worth mentioning that our preliminary experiments showed that the  \texttt{text-davinci-003} model was unable to generate a desired passage even with three examples in the prompt. We were able to generate good quality passages with ChatGPT, but it may be even more expensive once available through the API. Additionally, these models cannot perform a sequence of tasks step by step (e.g., curating a collection then queries, etc.). They may miss some parts of the sequence or do it all at once (generating documents and queries simultaneously), causing the automation of the training retrieval model to be difficult. 

\begin{algorithm}[t]
\caption{Our Synthetic Data Creation Approach}
\label{alg:greedy-cross}
\begin{algorithmic}[1]
\State \textbf{Input} (a) $T$: a target domain description; (b) $W$: a large, diverse, and heterogeneous collection (such as the Web); (c) $M_\theta$: a dense retrieval model  trained on the source domain; (d) $\widehat{M}$: an effective teacher model for pseudo labeling; (e) $N$: the desired size of synthetic collection ; (f) $k$: the iterative retrieval list size; (g) $k'$: the number of generated queries per document.
\State \textbf{Output} A dense retrieval model $M'$ for the target domain. 
\State $a_1, a_2, \cdots, a_{15} \leftarrow \textsc{DescriptionUnderstanding}(T)$
\State $q_\text{attr} \leftarrow \{a_1, a_2, \cdots, a_7\}$ 
\State $d_\text{attr} \leftarrow \{a_8, a_9, \cdots, a_{14}\}$
\State $r_\text{attr} \leftarrow \{a_{15}\}$
\State $S_{\text{seed}} \leftarrow \textsc{DocumentGen}(d_\text{attr})$
\State $C \leftarrow \emptyset$
\Repeat
    \State $d \leftarrow S_{\text{seed}}.pop()$
    \State $D \leftarrow \textsc{Retrieve}(\text{query} = d, \text{collection} = W, \text{count} = k)$
    \State $C \leftarrow C \cup D$
    \State $S_{\text{seed}} \leftarrow S_{\text{seed}} \cup D$
\Until{$|C| < N$}   
\State $Q \leftarrow \textsc{QueryGen}(C, q_\text{attr}, r_\text{attr}, k')$
\State $R \leftarrow \textsc{PseudoLabeling}(C, Q, r_\text{attr}, \widehat{M})$
\State $M' \leftarrow  \arg \min_{\theta} \mathcal{L}(M_{\theta}, \{Q, C, R\})$ \\
\Return $M'$
\end{algorithmic}
\end{algorithm}

To overcome all these obstacles, we propose an iterative document selection process (i.e., lines 7-14 in Algorithm \ref{alg:greedy-cross}). We first generate a document based on the domain attributes we extracted from the target domain description $T$. 
We call this generated document a seed document. 
We find that ChatGPT is the only language model that could successfully generate a related document given our document attributes. We tried T5, \textsc{T$k$-Instruct}, and GPT-3 and they could not generate a document with the given attributes. Instead, they generate a text using the words in the given instruction which is not 
sufficient for effective domain adaptation. 
We then run an iterative retrieval process using BM25 and a BERT-based cross-encoder reranking model trained on the source domain \cite{Nogueira:2019}. It retrieves $k$ documents (we empirically observe that $k$ should be set to a small value often less than $50$) in response to the seed document and then adds all the retrieved documents to the seed set. Again another document from the seed set is selected and another $k$ documents are retrieved. This process repeats until we reaches a collection $C$ with a desired synthetic collection size ($N$).

\subsubsection{\textbf{Synthetic Query Generation}}
In line 15 of Algorithm \ref{alg:greedy-cross}, we generate $k'$ queries per document in the constructed document collection $C$. To this aim, we train instruction-based $T5$ on MS MARCO for query generation using the MS MARCO query and relevance attributes. It is similar to the docT5query \cite{Nogueira2019FromDT}, but also takes query and relevance properties of the target domain as input. To be precise, we use form this input for the instruction-based $T5$ model: `Generate a query for the following Passage based on the given Attributes. Passage: $\cdots$. Attributes: $\cdots$.' We include the query and relevance attributes in the instruction. Therefore, it learns to generate queries with the given properties. The model is trained with a maximum likelihood objective as follows:
\begin{equation*}
     -\sum_k \log P(q_k| q_{i < k}, q_{\text{attr}}, r_{\text{attr}}),
\end{equation*}

\noindent
where $q_k$ is $k$\textsuperscript{th} output query token, $q_{\text{attr}}$ is the extracted values for query attributes in the taxonomy, and $r_{\text{attr}}$ is the extracted values for relevance attribute. We use beam search with the size of $k'$. 

\subsubsection{\textbf{Pseudo Labeling}}
Research on weak supervision by \cite{Dehghani:2017,Zamani:2018} showed that we can use existing retrieval models to annotate documents for a given query set and train student models based on the annotated data. More recently, this approach has been found effective in unsupervised domain adaptation \cite{wang:2021}. We use a cross-encoder re-ranking model based on BERT \cite{Nogueira:2019} that is trained on MS MARCO (our source domain) as a teacher model and annotate documents through soft labeling: the input includes the query, the relevance notion, and a document, and the output scores are used as labels. Let $D_q \subset C$ be a set of documents that should be annotated for query $q$ by the pseudo-labeler. We construct $D_q$ as follows:
\begin{itemize}[leftmargin=*]
    \item $D_q$ includes the document that $q$ was generated from.
    \item $D_q$ includes 25 random documents from the top 100 documents retrieved by BM25.\footnote{We empirically observe that taking 25 random samples from the top 100 documents leads to more robust performance compared to using the top 25 documents.} 
    \item $D_q$ includes 25 random documents from the top 100 documents retrieved by the dense retriever $M_\theta$. 
\end{itemize}

\subsection{Dense Retrieval Adaptation}
Given the constructed training set with pseudo-labels, we use the following listwise loss function for adapting the dense retrieval model $M_\theta$ to the target domain. We used Contriever \cite{izacard:2021} (an unsupervised dense retrieval model trained using contrastive learning) that is fine-tuned on MS MARCO as our $M_\theta$. 
Let $D_{q} \subset C$ be the set of documents annotated for query $q \in Q$ through pseudo-labeling. We use the following listwise loss function for each query $q$:
\begin{equation*}
    \sum_{d, d' \in D_{q}}  \mathbbm{1}\{y^T_q(d) > y^T_q(d')\} |\frac{1}{\pi_q(d)} - \frac{1}{\pi_q(d')}| \log ( 1+ e^{y^S_q(d') - y^S_q(d)}),
\end{equation*}
where $\pi_q(d)$ denotes the rank of document $d$ in the result list produced by the student dense retrieval model, and $y^T_q(d)$ and $y^S_q(d)$ respectively denote the scores produced by the teacher and the student models for the pair of query $q$ and document $d$. This knowledge distillation listwise loss function is inspired by LambdaRank \cite{Burges2010FromRT} and is also used by \citet{Zeng:2022:CLDRD} for dense retrieval distillation.

In addition, we take advantage of the other passages in the batch as in-batch negatives. Although in-batch negatives resemble randomly sampled negatives that can be distinguished easily from other documents, it is efficient since passage representations can be reused within the batch \cite{DPR}.

\begin{table*}[t]
    \centering
    \caption{Domain adaptation results in terms of NDCG@10 and Recall@100. Bold numbers indicate the highest value in each column (excluding Oracle). The superscript $^*$ denotes statistically significant improvements compared to all the baselines with respect to a two-tailed paired t-test with Bonferroni correction ($p\_value < 0.05$).}
    \vspace{-0.3cm}
    \resizebox{\textwidth}{!}{
    \begin{tabular}{lllllllllllll}\toprule
        \textbf{Model} & \multicolumn{2}{c}{\textbf{TREC COVID}} & \multicolumn{2}{c}{\textbf{FiQA}} & \multicolumn{2}{c}{\textbf{SciFact}} & \multicolumn{2}{c}{\textbf{ArguAna}} & \multicolumn{2}{c}{\textbf{Quora}} \\
        & \textbf{NDCG@10} & \textbf{R@100} & \textbf{NDCG@10} & \textbf{R@100} & \textbf{NDCG@10} & \textbf{R@100} & \textbf{NDCG@10} & \textbf{R@100} & \textbf{NDCG@10} & \textbf{R@100} \\\midrule
        BM25 & 0.688 & \textbf{0.498} & 0.253 &  0.539 & 0.690 & 0.908 &  0.471 & 0.942  & 0.807 & 0.973 \\
        ANCE & 0.652 & 0.457 & 0.295 & 0.581 & 0.511 & 0.816 & 0.418 & 0.934 & 0.852 & 0.987 \\
        SBERT & 0.477 & 0.072 & 0.257 & 0.542 & 0.537 & 0.846 & 0.425 & 0.945 & 0.855 & 0.988 \\
        Contriever & 0.273 & 0.172 & 0.245 & 0.562 & 0.649 & 0.926 & 0.379 & 0.901 & 0.835 & 0.987 \\
        Contriever-FT & 0.596 & 0.407 & 0.329 & 0.656 & 0.677 & 0.947 & 0.446 & 0.977 & 0.865 & 0.993 \\
        HyDE & 0.593 & 0.414 & 0.273 & 0.621 & 0.691 & \textbf{0.964} & 0.466 & \textbf{0.979} & - & - \\
        ANCE - Cond. Query & 0.640	&	0.459 &	0.294	&	0.575	&	0.518	&	0.813	&	0.406	&	0.932	&	0.843	&	0.980 \\
        Contriever-FT - Cond. Query & 0.596	&	0.409	&	0.336	&	0.652	&	0.667	&	0.949	&	0.445	&	0.966	&	0.866	&	0.980 \\
        \midrule
        \textbf{Ours} & \textbf{0.737$^*$} & 0.481 & \textbf{0.344$^*$} & \textbf{0.684$^*$} & \textbf{0.695$^*$} & 0.957 & \textbf{0.497$^*$} & 0.967 & \textbf{0.881$^*$} & \textbf{0.995} \\ \midrule
        Oracle & 0.752	&	0.515	&	0.368	&	0.699	&	0.744	&	0.970 &	0.529	&	0.973	&	0.885 & 0.984 \\
        CE Reranker & 0.757 & 0.498 & 0.347 & 0.539 & 0.688 & 0.908 & 0.311 & 0.942 & 0.825 & 0.973 \\
        \bottomrule
    \end{tabular}
    }
    \label{tab:results}
\end{table*}

\begin{table*}[t]
    \centering
    \caption{Ablation Study in terms of NDCG@10 and Recall@100. Bold numbers indicate the highest value in each column (excluding Oracle). The superscript $^\blacktriangledown$ denotes statistically significant performance degrade compared to our method (the first row of the table). Significance is identified using a two-tailed pair t-test with Bonferroni correction ($p\_value < 0.05$).}
    \vspace{-0.3cm}
    \resizebox{\textwidth}{!}{
    \begin{tabular}{lllllllllllll}\toprule
        \textbf{Model} & \multicolumn{2}{c}{\textbf{TREC COVID}} & \multicolumn{2}{c}{\textbf{FiQA}} & \multicolumn{2}{c}{\textbf{SciFact}} & \multicolumn{2}{c}{\textbf{ArguAna}} & \multicolumn{2}{c}{\textbf{Quora}} \\
        & \textbf{NDCG@10} & \textbf{R@100} & \textbf{NDCG@10} & \textbf{R@100} & \textbf{NDCG@10} & \textbf{R@100} & \textbf{NDCG@10} & \textbf{R@100} & \textbf{NDCG@10} & \textbf{R@100} \\\midrule
        Ours & \textbf{0.737} & \textbf{0.481} & \textbf{0.344} & \textbf{0.684} & \textbf{0.695} & \textbf{0.957} & \textbf{0.497} & \textbf{0.967} & \textbf{0.881} & \textbf{0.995} \\ 
        Ours w/o pseudo-labeling & 0.691$^\blacktriangledown$	&	0.473	&	0.336$^\blacktriangledown$	&	0.671$^\blacktriangledown$	&	0.687$^\blacktriangledown$	&	0.907$^\blacktriangledown$	&	0.477$^\blacktriangledown$	&	0.919$^\blacktriangledown$	&	0.852$^\blacktriangledown$	&	0.963$^\blacktriangledown$ \\
        Ours w/o seed document generation & 0.688$^\blacktriangledown$	&	0.399$^\blacktriangledown$	&	0.310$^\blacktriangledown$	&	0.660$^\blacktriangledown$	&	0.630$^\blacktriangledown$	&	0.874$^\blacktriangledown$	&	0.441$^\blacktriangledown$	&	0.882$^\blacktriangledown$	&	0.822$^\blacktriangledown$	&	0.919$^\blacktriangledown$ \\
        Ours w/o \emph{interactive} synthetic corpus creation & 0.704$^\blacktriangledown$	&	0.478	&	0.343	&	0.638$^\blacktriangledown$	&	0.662$^\blacktriangledown$	&	0.935$^\blacktriangledown$	&	0.481$^\blacktriangledown$	&	0.954	&	0.841$^\blacktriangledown$	&	0.993 \\
        \bottomrule
    \end{tabular}
    }
    \vspace{-0.3cm}
    \label{tab:ablation}
\end{table*}

\section{Experiments}
\label{sec:exp}
This section describes our datasets, experimental setup, and results. 

\subsection{Tasks and Data}
For evaluating our domain adaptation solution, 
we chose the target collections to be as diverse as possible within the public test collections in the BEIR benchmark \cite{thakur:2021}. Below we provide brief explanations of these collections. 

\label{sec:data}
\paragraph{\textbf{Source Domain}} As the source domain, we focus on passage retrieval provided by the MS MARCO collection \cite{MSMARCO}. As the standard practice on zero-shot learning offered by BEIR benchmark, most of baselines models have been pre-trained on this dataset, as our source domain. It contains 8.8 M passages and an official training set of 532,761 query-passage pairs collected from the Bing search log. Queries often have only one relevant passage per query, and  the relevant label is binary.


\paragraph{\textbf{Target Retrieval Task 1: Bio-Medical IR}} Our first target retrieval task focuses on retrieving scientific documents for biomedical queries. We use the collection provided by the TREC Covid Track in 2020 (\textbf{TREC-COVID}) \cite{voorhees:2020}, which is an ad-hoc retrieval task based on scientific documents related to the Covid-19 pandemic offered by the CORD-19 corpus \cite{lu:2020}. Similar to \citet{thakur:2021}, we use the July 16, 2020 version of CORD-19 collection as the target corpus, and the final cumulative judgments with query descriptions from the original task as test queries. The test collection consists of 50 test queries and a corpus of 171K documents.

\paragraph{\textbf{Target Retrieval Task 2: Financial Question Answering}} Our second task studies answer passage retrieval in response to natural language questions in the financial domain. We use the FiQA-2018 Task 2  \cite{maia:2018} (\textbf{FiQA}) that focused on answering questions based on personal opinions. The document collection was created by crawling posts on StackExchange under the Investment topic from 2009-2017, which serves as the corpus with 57K documents. The test set consists of 648 queries.

\paragraph{\textbf{Target Retrieval Task 3: Argument Retrieval}} This task explores ranking argumentative texts from a collection based on relevance to a given query on various subjects. We use the \textbf{ArguAna} dataset \cite{wachsmuth:2018} which has passage-level queries. The goal is to retrieve the most suitable counterargument for a given argument. The collection was collected from online debate portals. There are 1,406 argument queries in the dataset and the corpus size is 8.67K.

\paragraph{\textbf{Target Retrieval Task 4: Duplicate Question Retrieval}}: The aim of duplicate question retrieval is to detect repeated questions asked on community question-answering (CQA) forums. We use the \textbf{Quora} dataset that consists of 522,931 unique questions in corpus and 10,000 test queries.

\paragraph{\textbf{Target Retrieval Task 4: Fact Checking}} Fact checking involves verifying a statement against a large pool of evidence. It requires knowledge of the statement and the ability to analyze multiple documents. In a retrieval setting, the query is a claim, and we attempt to retrieve documents that confirms or refutes the claim. We use the \textbf{SciFact} collection \cite{wadden-etal-2020-fact} that consists of 300 scientific claims as test queries and 5K paper abstracts as the corpus.


\paragraph{\textbf{Constructing the heterogeneous Collection $W$:}} As explained in Section \ref{sec:formal}, $W$ is a heterogeneous collection of documents from which our model selects documents to synthesize the target retrieval corpus. To create this collection, we ensure that there is no document leakage between the target retrieval tasks and  $W$.\footnote{Note that document leakage is not necessary an issue in this task. In real world, the Web contains various types of documents that can satisfy the attributes of each target domain (e.g., each BEIR collection). The main challenge is to identify and recover these documents from a large heterogeneous corpus.} 
We create $W$ by putting together the documents from MS MARCO \cite{MSMARCO}, SciDocs \cite{Cohan:2020}, NFCorpus \cite{boteva:2016}, Touche-2020 \cite{Bondarenko:2021}, and CQADupStack \cite{Hoogeveen:2015}. This results in a collection with 9M+ documents.


\subsection{Experimental Setup and Evaluation Metrics}
We implemented and trained our models using TensorFlow. The network parameters were optimized using Adam \cite{AdamOptimizer} with  linear scheduling and the warmup of 4,000 steps. The learning rate was selected from $[\num{1e-6}, \num{1e-5}]$ with a step size of $\num{1e-6}$. The batch size was set to $128$. We set $k$ to 30, $N$ to 10,000, and $k'$ to 5 (see Algorithm \ref{alg:greedy-cross}). We use the BERT \cite{devlin:2018} with the pre-trained checkpoint made available from Contriever-FT \cite{izacard:2021} as the initialization.  Hyper-parameter selection (for both BM25 and neural models) and early stopping was conducted based on the performance in terms of MRR on the MS MARCO validation set. For query generation we use the T5 model from \cite{Nogueira2019FromDT}. As the re-ranking teacher model for pseudo labeling, we use a BERT cross-encoder, similar to \cite{Nogueira:2019}. For domain description understanding, we use three examples in the ChatGPT instruction.
Following BEIR \cite{thakur:2021}, we use NDCG@10 and Recall@100 as evaluation metrics. We use a two-tailed paired t-test for identifying statistically significant performance differences using Bonferroni correction with $p\_value < 0.05$.



\subsection{Results and Discussion}
We compare our method against the following baselines:
\begin{enumerate}[leftmargin=*]
        \item BM25 \cite{Robertson:1995}: an effective term matching retrieval method that evaluates and ranks a group of documents based on the presence of query terms regardless of their position in each document.
        \item ANCE \cite{xiong:2020}: a bi-encoder dense retrieval model that constructs hard negatives from an Approximate Nearest Neighbor (ANN) index of the corpus based on the model's representations. Consistent with previous works, we used RoBERTa \cite{Liu:2019} as the base language model that is trained on MS MARCO for 600K steps for our experiments.
        
        \item SBERT \cite{reimers:2019}: another dense retrieval baselines that uses BERT that employs Siamese and Triplet network architectures to generate sentence embeddings.
        
        \item Contriever \cite{izacard:2021}: an unsupervised dense retrieval model that learns adaptive representation via contrastive learning.

        \item Contriever-FT \cite{izacard:2021}: the Contriever model that is fine-tuned on MS MARCO training set. 
        
        \item HyDE \cite{gao:2022}: it utilizes GPT-3 to generate a hypothetical document. Then it uses Contriever to retrieve from the corpus with the hypothetical document as the query. This work has been proposed concurrent to our work. 
        
        \item ANCE - Cond Query: following \citet{asai:2022}, which is another concurrent work to ours, we concatenate the domain description with the query in ANCE so the query encoder is aware of the domain description.
        
        \item Contriever-FT - Cond Query: this is similar to the last baseline, but users Contriever-FT as the dense retrieval model.

    \end{enumerate}

As a source of reference we compare against the following approaches:(1) Oracle: this is our proposed approach that, instead of document collection construction, uses the target domain collection for query generation; and (2) CE Reranker: this is a BERT-based cross-encoder reranker trained on MS MARCO, which reranks the top 100 documents returned by BM25. Since this is not a dense retrieval model, we only report its results as a point of reference.

The results are reported in Table \ref{tab:results}.  We observe that dense retrieval baselines have difficulties surpassing the BM25 performance on TREC COVID, SciFact, and ArguAna datasets in terms of NDCG@10 in a zero-shot setting. This demonstrates the difficulty of dealing with distribution shift in neural information retrieval. HyDE that uses GPT-3 for generating hypothetical documents for test queries performs well in terms of Recall@100 on SciFact and ArguAna datasets. The proposed approach outperforms all dense retrieval baselines in terms of NDCG@10 in all collections. These improvements are statistically significant in all cases. It also better than its counterparts in terms of Recall@100 on  FiQA and Quora. Interestingly, our approach is the only dense retrieval model that can beat BM25 on TREC COVID and ArguAna. This demonstrates the effectiveness of our data creation pipeline. 

The performance gap between the Oracle model and the baselines is often less than 10\%, confirming the quality of the synthetic corpus our model creates. The Oracle model performs better than the proposed approach in all cases, except for Recall@100 on Quora. Note that the Oracle model does not necessarily provide upper-bound results, it just uses the target domain collection instead of synthetic collection construction. This results suggest that it is likely to construct a collection that dense retrieval models benefit from for adaptation, even more than the actual target collection. 
Our model outperforms the cross encoder reranker model in terms of Recall@100 in all cases, except for TREC COVID. 




\begin{figure*}[t]
    \centering
    \vspace{-0.3cm}
    \subfigure{
        \label{fig:sensitivity:a}
        \includegraphics[width=.32\textwidth]{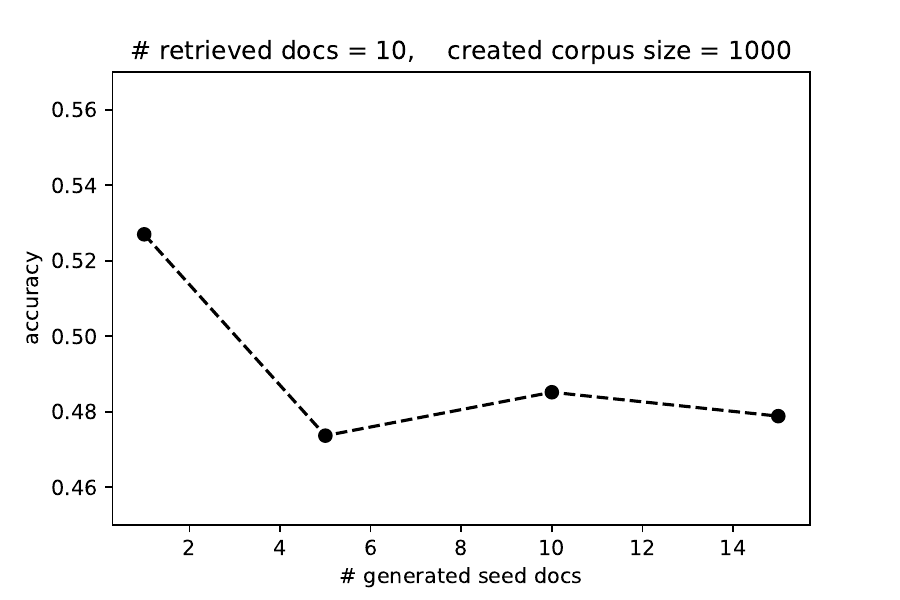}
    }
    \subfigure{
        \label{fig:sensitivity:b}
        \includegraphics[width=.32\textwidth]{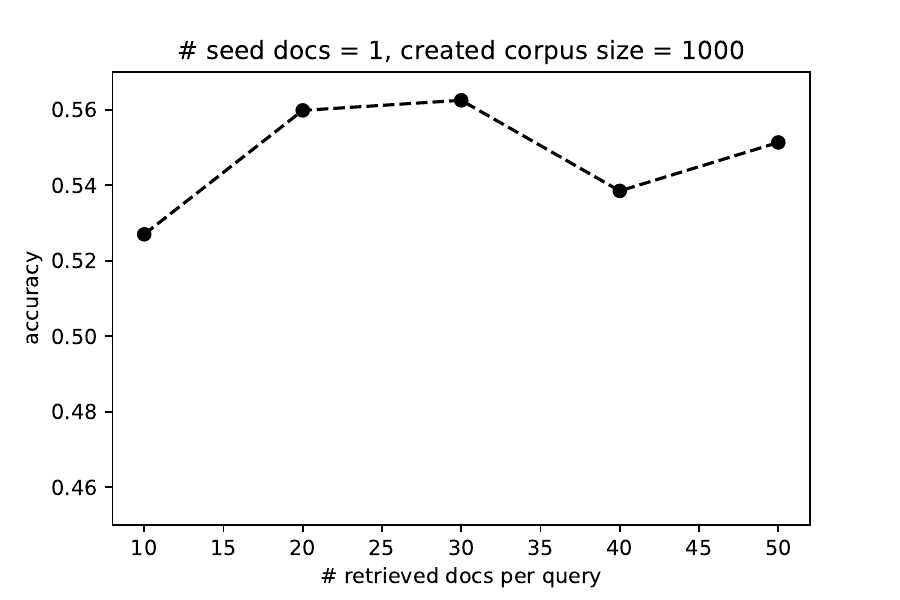}
    }
    \subfigure{
        \label{fig:sensitivity:c}
        \includegraphics[width=.32\textwidth]{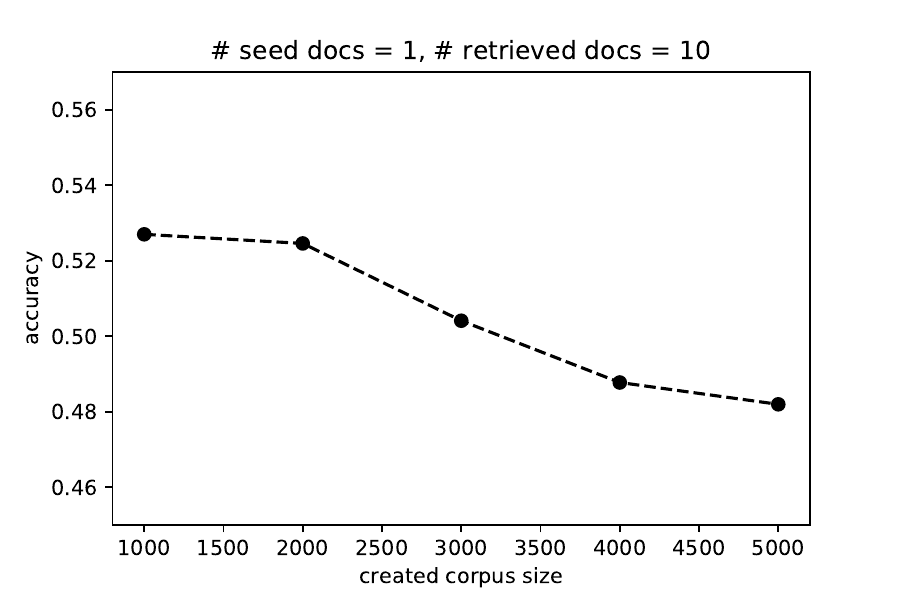}
    }
    \vspace{-0.4cm}
    \caption{Sensitivity of our iterative corpus creation process to different parameters in terms of average accuracy.}
    \vspace{-0.3cm}
    \label{fig:sensitivity}
\end{figure*}

\paragraph{\textbf{Ablation Study}}
To demonstrate the impact of each design decision we made in our pipeline, we ablate each major component in our model and report the results in Table~\ref{tab:ablation}. We first exclude the pseudo-labeling component (i.e., assuming that a document used for generating each query is relevant and any other document is non-relevant), and we observe statistically significant performance drop in nearly all cases. In the second ablation study, we exclude the seed document generation and use the domain instruction itself as the query to retrieve documents from $W$ and construct the collection $C$. This leads to even larger performance drop. Our last ablation focuses on converting the iterative collection construction part to a single retrieval run (i.e., retrieving 10,000 documents in response to the seed document). We observe that in this case, some collections hurt more than others. For example, performance drop on Quora is more significant than FiQA and TREC COVID. But generally speaking, the iterative process leads to a better performance. 

\paragraph{\textbf{Evaluating the Quality of the Synthetic Corpus Construction Approach}}
To provide a deeper look into the quality of the corpus that we construct in our model, we take the union of $W$ and all the target domain collections listed above. We then run our synthetic corpus construction experiment to see the accuracy of the model in retrieving the documents that actually belong to the target corpus. We report the average performance in Figure~\ref{fig:sensitivity}. In the left plot, we vary the number of generated seeds by ChatGPT and we observe that a single seed document is sufficient and including more documents degrades the accuracy of constructed collection. In the middle plot, we vary the number of retrieved documents per query (i.e., $k$ in Algorithm \ref{alg:greedy-cross}) and observe that the model shows a relatively stable performance compared to various values of $k$, however, smallest value led to the poorest performance.  In the last experiment, we increase the synthetic corpus size from 1,000 to 5,000 and observe that the accuracy of reconstructing document from the actual target domain decreases. However, this performance decrease is not substantial, and the accuracy is still higher than 48\% when selecting 5,000 documents. This is another signal to show that the proposed approach for corpus construction performs effectively.

\begin{table*}[t]
    \centering
    \caption{Retrieval description understanding results for each attribute in our taxonomy. We use ROUGE-L and Exact Match (EM) in addition to manual annotation to evaluate the model. Average results across 15 datasets are reported.}
    \vspace{-0.3cm}
    \resizebox{\textwidth}{!}{
    \begin{tabular}{clcccccccccccc}\toprule
        & & \multicolumn{3}{c}{Instruction Only} & \multicolumn{3}{c}{Instruction + 1 Example} & \multicolumn{3}{c}{Instruction + 2 Examples} & \multicolumn{3}{c}{Instruction + 3 Examples} \\
        & \textbf{Retrieval Attribute} & \textbf{ROUGE-L} & \textbf{EM} & \textbf{Manual} & \textbf{ROUGE-L} & \textbf{EM} & \textbf{Manual} & \textbf{ROUGE-L} & \textbf{EM} & \textbf{Manual} & \textbf{ROUGE-L} & \textbf{EM} & \textbf{Manual} \\\midrule
        \multirow{6}{*}{\rotatebox[origin=c]{90}{Query Attributes}} & Query topic & 0.800 & 0.800 & 0.800 & 0.711 & 0.666 & 0.733 & 0.733 & 0.733 & 0.733 & 0.733 & 0.733 & 0.733 \\
        & Query linguistic features  & 0.600 & 0.600 & 0.600 & 0.800 & 0.800 & 0.800 & 0.866 & 0.866 & 0.866  & 0.866 & 0.866 & 0.866 \\
        & Query language & 0.666 & 0.666 & 0.667 & 1.000 & 1.000 & 1.000 & 0.866 & 0.866 & 0.866 & 1.000 & 1.000 & 1.000 \\
        & Query structure & 0.099 & 0.066 & 0.133 & 0.866 & 0.866 & 0.800 & 0.933 & 0.933 & 0.933 & 1.000 & 1.000 & 1.000 \\
        & Query modality & 0.000 & 0.000 & 1.000 & 1.000 & 1.000 & 1.000 & 1.000 & 1.000 & 1.000 & 1.000 & 1.000 & 1.000 \\
        & Query format & 0.662 & 0.533 & 0.733 & 0.822 & 0.733 & 0.800 & 0.811 & 0.733 & 0.933 & 0.866 & 0.866 & 1.000\\ \hdashline
        \multirow{7}{*}{\rotatebox[origin=c]{90}{Doc Attributes}} & Document topic & 0.666 & 0.666 & 0.733 & 0.733 & 0.733 & 0.733 & 0.733 & 0.733 & 0.800 &  0.800 & 0.800 & 0.800 \\
        & Document linguistic features & 0.800 & 0.800 & 0.800 & 0.800 & 0.800 & 0.800 & 0.866 & 0.866 & 0.866 & 0.933 & 0.933 & 0.933\\
        & Document language & 0.266 & 0.266 & 0.266 & 0.800 & 0.800 & 0.800 & 0.800 & 0.800 & 0.800 & 0.866 & 0.866 & 0.866\\
        & Document structure  & 0.066 & 0.066 & 0.066 & 0.733 & 0.733 & 0.733 & 0.933 & 0.933 &  0.933 & 1.000 & 1.000 & 1.000 \\
        & Document modality & 0.000 & 0.000 & 1.000 & 1.000 & 1.000 & 1.000 & 1.000 & 1.000 & 1.000 & 1.000 & 1.000 & 1.000 \\
        & Document format & 0.377 & 0.200 & 0.533 & 0.677 & 0.600 & 0.866 & 0.800 & 0.800 & 0.867 & 0.711 & 0.666 & 0.800 \\
        & Document source & 0.836 & 0.800 & 0.866 & 0.826 & 0.533 & 0.866 & 0.893 & 0.666 & 0.933 & 0.933 & 0.733 & 0.933 \\ \hdashline
        & Relevance notion  & 0.524 & 0.133 & 0.466 & 0.689 & 0.533 & 0.800 & 0.701 & 0.533 & 0.666 & 0.807 & 0.733 & 0.866 \\ \midrule
        & \textbf{Average} & 0.454 & 0.400 & 0.619 & 0.818 & 0.771 & 0.843 & 0.852 & 0.819 & 0.871 & 0.894 & 0.871 & 0.914 \\ 
         \bottomrule
    \end{tabular}
    }
    \vspace{-0.3cm}
    \label{tab:desc_understanding}
\end{table*}

\paragraph{\textbf{Analyzing the Domain Description Understanding Component}}

As described in Section \ref{sec:desc-und}, we provided three IR experts (not the authors of this work) with all 15 public collections in the BEIR benchmark, and asked them to come up with a description for each retrieval task associated with each collection in a collaborative session. We later presented them with the our taxonomy and asked them to annotate the descriptions accordingly. The input of the description understanding model is the task description, in addition to arbitrary choice of examples, and the output is expected to be the value of taxonomy attributes.

Considering we cast the problem of description understanding to a sequence-to-sequence format, following the literature, we used ROUGE-L \cite{lin-2004-rouge} and Exact Match as our evaluation metrics. ROUGE-L (Recall-Oriented Understudy for Gisting Evaluation) is a commonly used evaluation metric in NLP for summarization tasks, measuring overlap between n-grams in reference summaries and the generated summary. The "L" refers to the longest common subsequence. ROUGE-L scores range from 0 to 1, with 1 being a perfect match. Exact Match (EM) measures the percentage of predictions that exactly match the ground truth, with 1 being a perfect match and 0 no match. Since the task is generative, automatic metrics may not be sufficient, so three annotators manually labeled the outputs of each model, scoring 1 if desirable and 0 if not. Final labels were decided through majority voting.

Table \ref{tab:desc_understanding} presents the results of ChatGPT for domain description understanding. We made sure that the model is not benefiting from any session data, by initiating a new session for each experiment. Each cell displays the average of scores for a particular attribute across 15 collections. The last row reflects the overall performance of each setting based on the average of all attributes. As expected, the highest performance is mostly achieved with Instruction and 3 examples is given. The reason is that the model receives more examples, thus has a better chance of encountering similar cases. As Table \ref{tab:desc_understanding} illustrates, the results of the manual metric highly correlate with the automatic metrics, except for the query and document modality attributes in the instruction only setting. We observe that in this setting, modality attributes resulted in 0.00 with the automatic metrics, but they resulted in 1 in manual annotation. After looking into results, we figured the disparity is because ground truth labels the modality feature as uni-modal, multi-modal, etc. but the sequence-to-sequence model labels it differently, e.g., text. This issue resolves after seeing one example in prompt. We also observe that query and document structure attributes result in a close-to-zero performance in the instruction-only setting. This may be due to the fact that in our instruction, we only provided the model with examples of values for these attributes. However, these attributes have been implicitly mentioned in the domain descriptions, and some in-domain knowledge is necessary to interpret the structure or modality of the task. Again, the performance would significantly improve after seeing only one example. Note that all datasets within BEIR are  unstructured, so the model may repeat the only label it has given as example for structure and modality attributes.

Further, we observe that relevance notion is one of the hardest attributes to predict. This makes sense because usually, understanding what constitutes relevance requires a deep understanding of the task, which these models currently lack. A deep dive into the results showed us that in many cases, the model generalizes the query attributes to the document attributes, especially in cases that are not explicitly described. For example, if the query topic attribute was predicted as ``medical,'' the model may generalize it to the document topic as well. However, we know that IR features are not necessarily symmetric. A medical query could request information from a heterogeneous corpus such as the Web, and the symmetric assumption makes data synthesis unrealistic.


\vspace{-0.3cm}
\section{Conclusions and Future Work}

This paper introduced a new category of domain adaptation methods for neural information retrieval and proposed a pipeline that leverages target domain descriptions to construct a synthetic target collection, generate queries, and produce pseudo-relevant labels. The results of experiments conducted on five diverse target collections demonstrated that our proposed approach outperforms existing dense retrieval baselines in such a domain adaptation scenario. 
This work holds the potential for practical applications where the target collection and its relevance labels are unavailable, while preserving privacy and complying with legal restrictions. Future work involves incorporating additional domain-specific information, such as data source and language, and evaluating its conceptualizing ability with more implicit descriptions.

\medskip

{ \noindent {\textbf{Acknowledgements.}}
This work was supported in part by the Center for Intelligent Information Retrieval and in part by a Bloomberg Data Science PhD Fellowship. Any opinions, findings and conclusions or recommendations expressed in this material are those of the authors and do not necessarily reflect those of the sponsor.}

\bibliographystyle{ACM-Reference-Format}
\bibliography{acmart}

\end{document}